# Statistically constrained operator associated with additivity of communication channel

Toshio Fukumi*
Matsuyama University
4-2 Bunkyo, Matsuyama City, Ehime 790 8578, Japan

**ABSTRACT**

The power of quantum communication channel arises from additively of Holevo information, which certainly depends on the fact that the Hilbert space grows exponentially. The additively will be be shown to be achieved by stochastically maximizing a Hilbert space.   In the conjecture we introduce operator statistics which is a natural extension of functions of operator, and employing it we discuss that the information is additive in probability.

## 1. INTRODUCTION

  The capacity of quantum information channel is a current topic in the quantum information science, in which most interesting problem is whether quantum communication channel will be able to overcome the capacity of classical one. Recently, Giovannetti and Lloyd discussed additivity properties of Gaussian channel.[1] Let me refer Giovannetti and Lloyd who wrote " One of the most challenging open questions of quantum communication theory is the additivity of the various quantities charactering the information transmission in a chanel.[2] The issue at hand is whether quantum entanglement is able to improve the performance of classical protocol.[3,4] The supposed aditivity of the Holevo information[5] is the most important example of this kind of issue. " and " The additivitiy of the Holevo information has been linked to the additivity of other quantities. In particular, it is known that proving the additivity of the Holevo information is equivalent to proving the additivity of the minimum von Neumann entropy at the output of the channel ". One of the possible difficulties of their arguments, however, will be that Gaussian statistics is based on the central limit theorem in which elements should be independent large number. As the domain of definition of the Heisenberg operator of entanglement is not dense any where in a Hilbert space which means that operator does not have limit in it's domain of definition nor we cannot talk about norm convergence. Therefore we will not be able to expect Gaussian statistics. Actually entanglement is a rare event.[6]   In this situation, Poisson statistics will be most probable candidate to describe the behavior of quantum entanglement. Hereafter we will travel along this avenue. Firstly, we decompose Heisenberg operator into spectral measure whose kernel is Poisson. Secondly, we introduce operator valued Poisson process where Poisson process stands for a rare event. Thirdly, operator valued Poisson process is converted into projection valued Poisson process. Finally, employing the independence of sample path, coarse  $\sigma$ -additivity of Holevo information will be derived, that is, Holevo information is additive in probability in this case. Underlying mathematics of present attempt is to prepare a larger Hilbert space in order to incorporate additivity of quantum quantities. Indeed we should prepare a larger Hilbert space to describe entanglement than that for many body coherent state. To be specific, many-body coherent state can be in a compact Hilbert space, whereas entanglement cannot. Present attempt is to maximize a Hilbert space stochastically in order to discuss quantum quantities in terms of probability theoretic point of view. In particular we employ operator valued Poisson stochastic process and projection valued Poisson random process in which any convergence or limit is described in probability. Along the way, we employ spectral decomposition of unitary operator into Poisson random process and then reconstruct the quantum geometrical object via random tessellation. It will be shown that there exist a bounded linear projection operator associated the probability distribution of unitary operator which is additive in probability. Additivity in weak topology will be discussed in the appendix.

## 2. Unitary operator

An operator is unitary if it is invertible and
$$U^{-1} = U^*$$
holds. Unitary operator can be derived from Heisenberg operator by Cayley transform
$$U_H = \frac{H - iH}{H + iH}.$$

## 3. Spectral decomposition

The present purpose is to employ spectral decomposition of unitary operator into Poisson random process, and reconstruct a quantum geometrical object via random tessellation.

### 3.1 Spectral decomposition

Spectral decomposition of self-ad joint operator is given by
$$H = \int_{-\infty}^{\infty} \lambda dE(\lambda),$$
*where*
$$\{E(\lambda)\} : resolution\ of\ unity.$$

### 3.2 Measure transformation

The measure transformation associated with Cayley transformation is given by
$$H = \int \theta dE(\theta),$$
$$H = H^*$$
$$U = \int_0^{2\pi} e^{i\theta} dE(\theta),$$
$$U^{-1} = U^*.$$

## 4. Poisson process

Gaussian process is describes an ensemble of large number whereas Poisson process is responsible to rare event. In this study we employ Poisson process whose kernel is given by
$$P(U)(t) = \sum_{n \in Z} \frac{U^n}{n!} e^{-U}(t).$$

Then we have the spectral measure[7]

$$(P(U)h \mid h) = \int_T P(\theta) d\mu_h(\theta).$$

## 5. Projection valued measure

There exists a bounded linear operator such that
$$(\Phi(f)h_1 \mid h_2) = \int_T f d\gamma_1 d\gamma_2.$$
This is a projection operator. That is
$$\Phi^2 = \Phi$$
and orthogonal
$$\Phi^* = \Phi.$$
In the present case we find
$$\Phi(P) = P(U).$$
This projection valued measure[8] defined in a Borel set and is a completely positive map which defines Holevo information. It will be shown that this projection valued measure is additive in probability.

## 6. Additivity of channel

Since the projection valued operator is in a positive independent set in a Borel set we can conclude that Holevo information is coarse σ-additive. This means Holevo information is additive almost surely. We are employing an ensemble of product state
$$H_1 \otimes H_2 \otimes \cdots \otimes H_n.$$
Nonetheless we can encode information as follows.

## 7. Coding and decoding

Let us consider how to code a geometrical object in a Hilbert space into an ensemble of entangled photons. For this purpose a stochastic differential equation(SDE)[9,10,11] provides a powerful tool. SDE is a differential equation in which inhomogeneous term is driven random motion. Following is an application of SDE. For a given function $X$ we have
$$dX(t) = -\text{grad}\, S(t) X(t) + \sqrt{\omega}\, \partial_t^* \Phi(t) dt,$$
$$\text{grad}\, S(t) = (S \oplus A) - (S \mid entanglement \mid A),$$
$$S : geometrical\ object\ in\ L^2,$$
$$A : a\ set\ of\ elementaly\ gates,$$
$$\omega : group\ velocity.$$
This equation is similar to that of Fukumi[12] who formulated annihilation and creation operators in stochastic mechanics, We can achieve decoding via random tessellation. Now the question is when it will stop. Following conditions are necessary.
1. Poisson process should be stationary.
2. Manifold should be isolated from segment. By segment we mean a set of elementary gates. That is
$$\text{grad}\, S(t) = 0.$$
3. Intersection should be zero.

4. *S* and *A* are mutually independent.

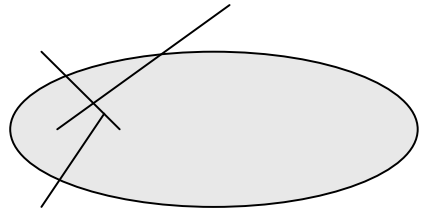

Fig.(1) Encoding via random hitting.

The decoding can be performed via random tessellation in which a random mosaic will be produced. The random mosaic contained information of geometrical object encoded by stochastic hitting. The decoding process is going the shortest way that is geodesic on manifold as illustrated in Fig.(2). This is random walk along geodesic, which is the most probable path in the present case.

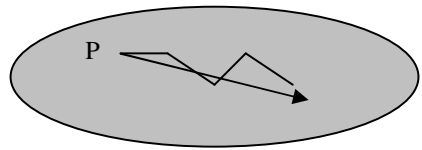

Fig.(2) Decoding process via random walk along geodesic on manifold.

and generates entanglement stochastically. Fig.(1) illustrates the hitting of quantum gates stochastically onto a target manifold which produces the deformation of the manifold.

## 8. Summary

Entanglement is not densely defined any where in a Hilbert space. This is the reason why we have a difficulty in discussing additivity of information in terms of metric. Present attempt is to employ deformation of Hilbert space $L^2(M_t)$ in which $M_t$ is a stochastically time dependent manifold and the metric is stochastic, where quantum quantities are also interpreted as stochastic variables. Present scenario enables us to treat additivity via stochastic manner. Thus we are maximizing a Hilbert space stochastically. Nonetheless we can encode and decode information by random process. Indeed we must prepare a larger space when we encounter a difficulty. Another approach will appear in my appendix.

## 9. Appendix

Entanglement is not dense in a Hilbert space. This is the main difficulty. But in what follow we try it to study in a weak topology. Let *H* be a Heisenberg operator for entanglement, then

(1) $H^*$ is closed,

(2) $H$ is closable iff $H^*$ is densely defined and $\bar{H} = H^{**}$.

Then we require
$$H \subseteq H^*$$
as a symmetric condition. Suppose $K$ is a dense linear subspace and omit asterisk except for explicitly noted, then we can derive square root of operator. For this purpose we expand space $K$ to be
$$K_H \subseteq K_{H_{1/2}}.$$
Then we define a square root of operator as
$$\left\| H^{1/2} f \right\|^2 = \langle f, Hf \rangle \leq \|f\| \|Hf\|.$$
Equation of motion is given by
$$\frac{\partial}{\partial t} = e^{-\sqrt{\omega} H^{1/2}}.$$
Actually this is a diffusion equation and a primitive form of Dirac equation and so it can be written by SDE form[10] as
$$dX(t) = \sqrt{\omega} \partial_t^* X(t) dt.$$

In this equation
$$\partial_t^* = \frac{\partial}{\partial \dot{B}(t)},$$

$$\dot{B}(t) = \frac{dB(t)}{dt} \quad : quite\ formal,$$

$$B(t) : Brownian\ motion.$$

This is possible as the domain of definition is dense and so we can take into account Brownian motion to describe diffusion process. In terms of Brownian motion we can formulate additivity in Gaussian measure in weak topology. Let us denote Dirac operator as $D$, then we find with orthonormal basis
$$D = \sum_{i=1}^{d} x(X_i) \nabla_{X_i}.$$
Let fix a point on the manifold as an origin and let $X = \{X_i\}$ be a moving frame obtained by parallel translating the frame along the geodesic rays. This moving frame has the property that all Christoffel symbols at the origin, and because the connection is torsion free, all the brackets $[X_i, X_j]$ vanishes as well. Hence force we can apply a SDE for random tessellation given by
$$dX(t) = -grad\, S(t) X(t) dt + \sqrt{\omega} \partial^* X(t) dt.$$
This equation can be integrated annalistically to be
$$X(t) = e^{\int_0^t grad\, S(t) ds - \frac{\omega - grad\, S(t)}{2} B(t)}.$$
This is a representation of geometrical Brownian motion.

*E-mail; tofukumi@cc.Matsuyama-u.ac.jp; Phone 81 089 925 7111